\documentclass[aps,prl,twocolumn,superscriptaddress,showpacs,showkeys,
notitlepage]{revtex4-1}

\usepackage{times,amsmath,amsfonts,amssymb,mathrsfs,graphics,graphicx,color,
comment,bm}
\usepackage[next]{inputenc}
\usepackage[dvips]{epsfig}
\usepackage[pdfstartview=FitH]{hyperref}
\usepackage{natbib}
\usepackage{pdfsync}
\usepackage{appendix}

\begin{document}
\title{$\mu$SR and magnetometry study of superconducting 5\% Pt doped IrTe$_2$}

\author{M.N. Wilson}
\address{Department of Physics and Astronomy, McMaster University, Hamilton,
Ontario L8S 4M1, Canada}
\author{T. Medina}
\address{Department of Physics and Astronomy, McMaster University, Hamilton,
Ontario L8S 4M1, Canada}
\author{T.J. Munsie}
\address{Department of Physics and Astronomy, McMaster University, Hamilton,
Ontario L8S 4M1, Canada}
\author{S.C. Cheung}
\address{Department of Physics, Columbia University, New York, New York 10027,
USA}
\author{B.A. Frandsen}
\address{Department of Physics, Columbia University, New York, New York 10027,
USA}
\author{L. Liu}
\address{Department of Physics, Columbia University, New York, New York 10027,
 USA}
\author{J. Yan}
\address{Department of Materials Science and Engineering,
University of Tennessee, Knoxville, Tennessee 37996, USA}
\address{Materials Science and Technology Division, Oak Ridge National
 Laboratory, Oak Ridge, Tennessee 37831, USA}
\author{D. Mandrus}
\address{Department of Materials Science and Engineering,
University of Tennessee, Knoxville, Tennessee 37996, USA}
\address{Materials Science and Technology Division, Oak Ridge National
 Laboratory, Oak Ridge, Tennessee 37831, USA}
\author{Y.J. Uemura}
\address{Department of Physics, Columbia University, New York, New York
 10027, USA}
\author{G.M. Luke}
\address{Department of Physics and Astronomy, McMaster University, Hamilton,
 Ontario L8S 4M1, Canada}
\address{Canadian Institute for Advanced Research, Toronto, Ontario M5G 1Z7,
 Canada}
\begin{abstract}
We present magnetometry and muon spin rotation ($\mu$SR) measurements
 of the superconducting dichalcogenide Ir$_{0.95}$Pt$_{0.05}$Te$_2$. From
 both sets of measurements we calculate the penetration depth and thence
 superfluid density as a function of temperature. The temperature dependence
 of the superfluid densities from both sets of data indicate fully gapped
 superconductivity that can be fit to a conventional s-wave model and yield
 fitting parameters consistent with a BCS weak coupling superconductor. We
 therefore see no evidence for exotic superconductivity in 
Ir$_{0.95}$Pt$_{0.05}$Te$_2$.
\end{abstract}

\maketitle
\section{Introduction}
Transition metal dichalchogenides have been studied for many years in an effort
 to understand their diverse properties \cite{Wilson1969,Rossnagel2011}. These
 materials are layered quasi-two dimensional systems that frequently exhibit
 charge density wave (CDW) ordering that is not yet fully understood
 \cite{Rossnagel2011}. Furthermore, the crystal structure of these materials is
 amenable to substitution and intercalation of a wide variety of dopant atoms to
 allow tuning through a broad range of electronic properties \cite{Friend1987}.
 In particular, these systems provide a valuable avenue to study the interplay
 of structural transitions and superconductivity as in many cases
 superconductivity emerges after the CDW transition is suppressed by
 doping or applied pressure 
\cite{Yokoya2001, Morosan2006, Yang2012, Sipos2008, Kusmartseva2009}.

IrTe$_2$ is a member of this group of compounds. It undergoes a structural
 transition at about 270~K \cite{Matsumoto1999} from the trigonal
 P$\overline{3}$m1 space group to triclinic P$\overline{1}$
 \cite{Cao2013, Pascut2014, Toriyama2014}. Recent work has shown
 that this structural transition is associated with a charge density wave that
 has a periodicity 6 times larger than the underlying lattice
 \cite{Ruan2015, Hsu2013, Li2014}. Substituting Ir with Pd, Pt, or Rh
 \cite{Pyon2012, Yang2012, Kudo2013, Ootsuki2012} or intercalation with
 Cu \cite{Kamitani2013} suppresses the structural transition and leads to
 superconductivity with a maximum $T_C$ of 3~K and $H_{C2}
 \approx$ 0.1~T. Intercalation with other transition metals also suppresses
 the structural transition but does not lead to superconductivity, possibly
 as a result of competing magnetism \cite{Yan2013}. Measurements of
 $T_C$ as a function of hydrostatic pressure in Pt-substituted IrTe$_2$
 have shown that increasing the temperature of the structural transition
 decreases $T_C$, which shows that the appearance of superconductivity
 is directly related to the disappearance of the structural transition
 \cite{Kiswandhi2013}.

IrTe$_2$ is of particular interest as both Ir and Te have high atomic
 numbers. Spin orbit coupling is therefore expected to be high which
 may lead to exotic states such as topological superconductivity
 \cite{Schnyder2008, Fu2010}. Determining the superconducting
 symmetry is important as unconventional (non s-wave) symmetry
 is required for superconductors to be topologically nontrivial \cite{Fu2010}.

Previous measurements of the superconducting symmetry by thermal
 conductivity \cite{Zhou2013} and STM \cite{Yu2014} suggest conventional
 s-wave superconductivity. However, the thermal conductivity measurements
 cannot conclusively rule out odd-parity p-wave superconductivity, and STM
 measurements are inherently a surface technique and so the state they
 probe may not be representative of the bulk superconductivity. Furthermore,
 no penetration depth measurements have been conducted on this material.
 These measurements are important, as the temperature dependence of the
 penetration depth gives information about the symmetry of the superconducting
 gap \cite{Sonier2000}.

Muon spin rotation ($\mu$SR) is a powerful technique that can be used
 to study the magnetic penetration depth of type II superconductors in
 the vortex state \cite{Sonier2000}. In this technique spin-polarized
 muons are implanted up to a few hundred $\mu$m into the sample
 where they precess in the local magnetic field and decay, emitting
 positrons that are detected to gain information about the local
 magnetic field. Importantly, the muons are implanted far enough
 into the sample that this can be considered a truly bulk technique.
 Therefore, surface effects that may change the states measured
 by techniques such as STM will not be a factor in these measurements. 

In this paper we present complementary  $\mu$SR and SQUID magnetometry
 measurements of the penetration depth of Ir$_{0.95}$Pt$_{0.05}$Te$_2$.
 These measurements indicate an s-wave superconducting state, with gap
 and $T_C$ values that are consistent with a conventional BCS weak-coupling
 superconductor.

\section{Experimental Methods}
Single crystals of Ir$_{0.95}$Pt$_{0.05}$Te$_2$ with sizes of a couple
 mm$^3$ were grown using the self flux growth method \cite{Fang2013}. 

Muon spin rotation ($\mu$SR) experiments were performed at the
 TRIUMF laboratory in Vancouver, Canada. We used the Pandora
 dilution refrigerator spectrometer on the M15 surface-muon beam line.
 This instrument gives access to temperatures between 0.03~K and 10~K
 with the sample mounted on a silver cold finger, magnetic fields up to
 5~T with a superconducting magnet, and a time resolution of 0.4~ns.
 The field is applied parallel to the incoming muon beam direction,
 and we performed measurements with the muon spin rotated
 perpendicular to the field direction (SR). These experiments were
 performed on an unaligned collection of small (\textless 1-2 mm)
 irregularly shaped single crystals mounted on a 1 x 2~cm silver
 plate using  Apiezon N-grease. We used the $\mu$SRfit software
 package to analyze the $\mu$SR data \cite{musrfit}

Magnetometry measurements were performed at McMaster University
 using a Quantum Design XL-5 MPMS with an iHelium He$^3$ cryostat
 insert for measurements down to 0.5~K. Magnetization vs. temperature
 curves were measured both on a subset of unaligned crystals from the
 $\mu$SR sample weighing 238~mg (polycrystalline sample), and on an
 aligned single crystal plate weighing  4.72~mg with dimensions 2.4~mm
 x 1.5~mm x 0.35~mm (C-axis). Magnetization vs. field curves were
 measured with fields up to 0.15~T and temperatures ranging from
 0.5 to 3~K using the single crystal plate. Alignment of the single crystal
 was verified with Laue X-Ray diffraction prior to the magnetometry
 measurements.

\section{Results and Discussion}

\begin{figure}[ht]
\includegraphics[width=\columnwidth]{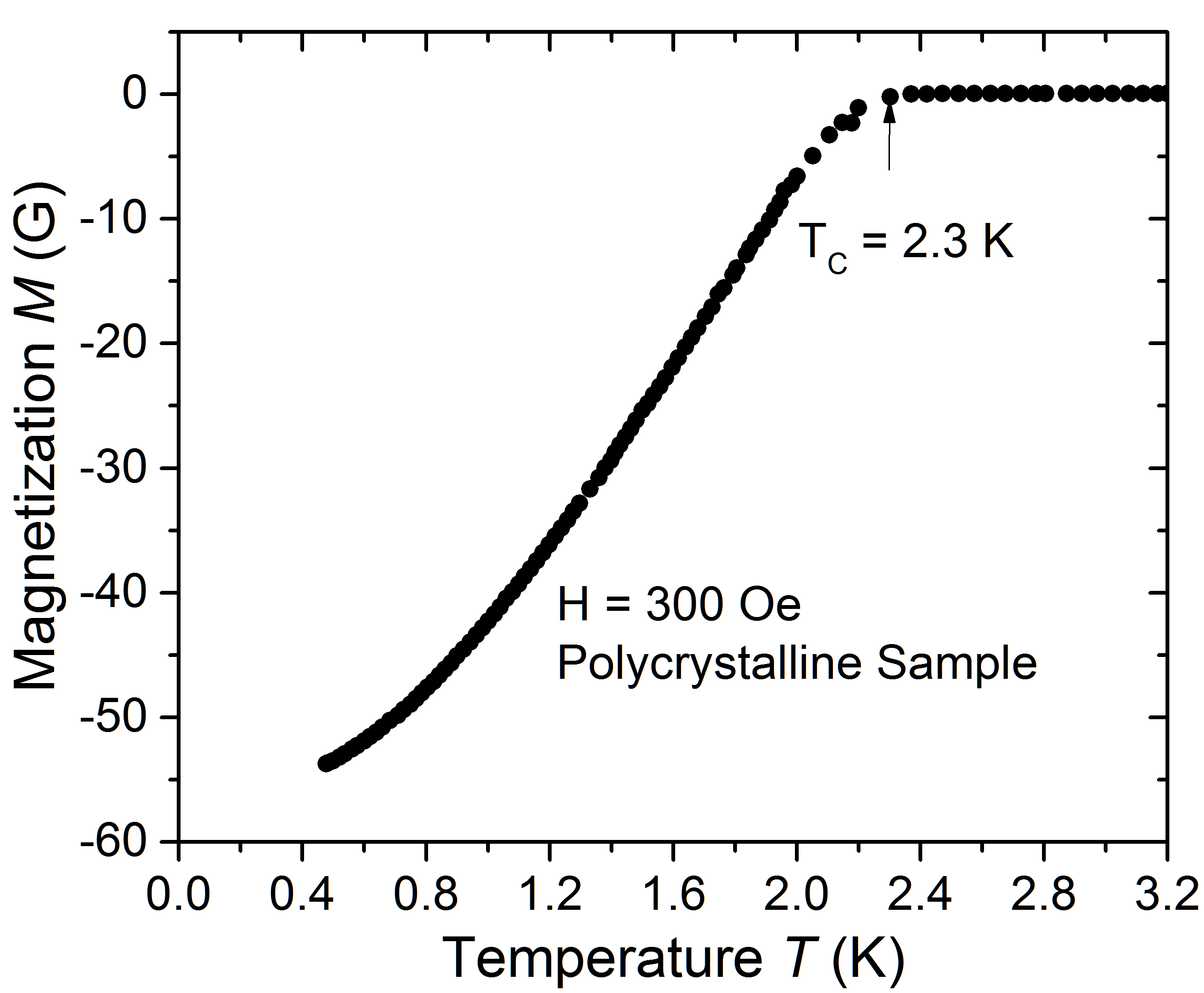}
\caption{Magnetization measurements on a polycrystalline sample of
 Ir$_{0.95}$Pt$_{0.05}$Te$_2$ measured in a field of 300~Oe after
 cooling in zero field.}
\label{fig:polyMvT}
\end{figure}

Figure \ref{fig:polyMvT} shows a temperature scan of the magnetization
 taken with an applied field of 300~Oe after cooling in zero field on the
 polycrystalline sample for comparison with the $\mu$SR data. This data
 shows strong diamagnetism, indicating that our sample is superconducting
 with a $T_c$ of about 2.3~K at $H_{ext} = 300$~Oe. 

\begin{figure}[ht]
\includegraphics[width=\columnwidth]{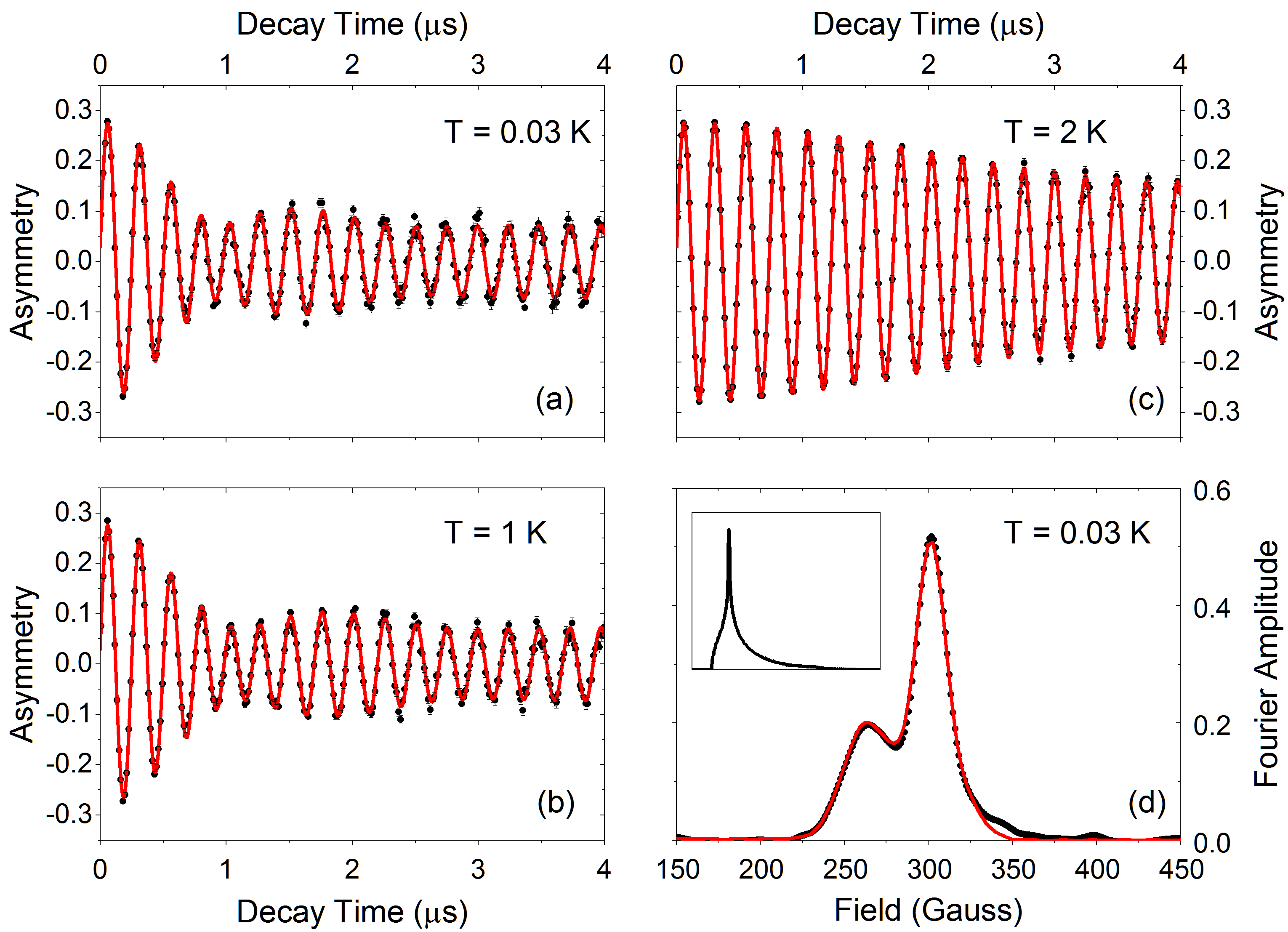}
\caption{SR $\mu$SR time spectra of Ir$_{0.95}$Pt$_{0.05}$Te$_2$ measured
 in an applied field of 300~Oe at (a) T=0.03~K, (b) T=1~K, and (c) T=2~K. (d)
 Fourier transform of the $\mu$SR data collected in an applied field of 300~Oe
 at T=0.03~K. The inset in (d) shows the theoretical field distribution of a
 superconductor using the London model \cite{Brandt1988} .}
\label{fig:TFSpectra}
\end{figure}

Figures \ref{fig:TFSpectra} (a-c) show $\mu$SR time spectra measured
 in an applied external field of 300~Oe \textless ~$H_{C2}$ transverse
 to the muon spins at 0.03~K, 1~K, and 2~K after field cooling the sample
 to ensure a uniform vortex lattice. This data shows a relaxing oscillating
 signal, with a beat evident in the lower temperatures along with a non-relaxing
 signal that persists to large times. This indicates the presence of more than
 one component to the signal, and can be more easily visualized by looking
 at the Fourier transform (FT) of the 0.03~K data found in Fig.
 \ref{fig:TFSpectra}~(d). We interpret the two peaks in the FT as arising
 from muons missing the sample and landing in the silver sample holder
 (peak at $\approx$ 300~G) and those hitting the sample and probing the
 superconducting state (lower field peak). 

Muons that land in a superconducting sample with an applied field between
 $H_{C1}$ and $H_{C2}$ see an asymmetric field distribution arising from
 the vortex state that will have the form shown in Fig. \ref{fig:TFSpectra}
 (d) inset. The experimental data from such a measurement, even on an
 ideal vortex lattice, will always show some broadening of this distribution
 due to the finite lifespan of the muon and time-window of the experiment.
 In practice, inhomogeneities in a sample will cause additional broadening
 of the field distribution that is difficult to rigorously account for. This is
 particularly important for the case of a polycrystalline sample where
 varied orientation and possible slight differences between the properties
 of different grains will broaden the signal. For our sample, we fit the field
 distribution to a three component model shown in Eq. \ref{eq:musr}
 similar to that used by Khasanov et al. in measurements on high $T_C$
 cuprates \cite{Khasanov2007}. This fit has two Gaussian-relaxing
 components representing the asymmetric superconducting line shape,
 and one non-relaxing component representing the silver background.
 These fits are made in the time domain to avoid Fourier transform
 broadening and to properly use the experimental error bars for weighting.

\begin{equation}
\begin{split}
A = & A_T \left[ C\left( F \cos(\gamma_{\mu}B_1 t) + (1-F)
 \cos(\gamma_{\mu}B_1  t)e^{-0.5(\sigma_1 t)^2} \right) \right.\\
& \left .+  (1 - C)\left(\cos(\gamma_{\mu}B_2  t)
e^{-0.5(\sigma_2 t)^2}\right) \right]
\end{split}
\label{eq:musr}
\end{equation}

Here, $C$ and $F$ are temperature independent values giving the ratio of the
 three components, $B_1$ is the temperature independent mean field for the
 silver site, $B_2$ is the temperature dependent mean sample field, and
 $\sigma_i$ are the temperature dependent Gaussian relaxation rates.

In this case the penetration depth can be determined from the equation
 \cite{Brandt1988}:

\begin{equation}
\lambda = \sqrt{\frac{0.043\sqrt{2} \gamma_{\mu} \phi_0}{\sqrt{\langle
 (\Delta F)^2\rangle}}}.
\label{eq:lambda}
\end{equation}
Here, $\gamma_{\mu} = 135.538$~ MHz/T is the muon gyromagnetic ratio,
 $\phi_0 = 2.06783$~Wb is the flux quantum, and $\langle (\Delta F)^2\rangle$
 is the central second moment (variance) of the fit frequency distribution.
 $\langle (\Delta F)^2\rangle$ is given by Eq. \ref{eq:sigma}, which can be
 derived by considering that the second moment of a sum of two Gaussian
 distributions is the sum of the individual second moments, and that the
 central second moment is the second moment minus the square of the
 mean \cite{Cramer1946}.

\begin{equation}
\langle (\Delta F)^2\rangle = R_1\sigma_1^2+R_2\sigma_2^2 +
 R_1R_2(\gamma_{\mu}B_1 - \gamma_{\mu}B_2)^2
\label{eq:sigma}
\end{equation}
Here, $\sigma_i$ are the relaxation rates, $B_i$ are the mean fields,
 and $R_i$ are the relative weights of the two components.

\begin{figure}[ht]
\includegraphics[width=\columnwidth]{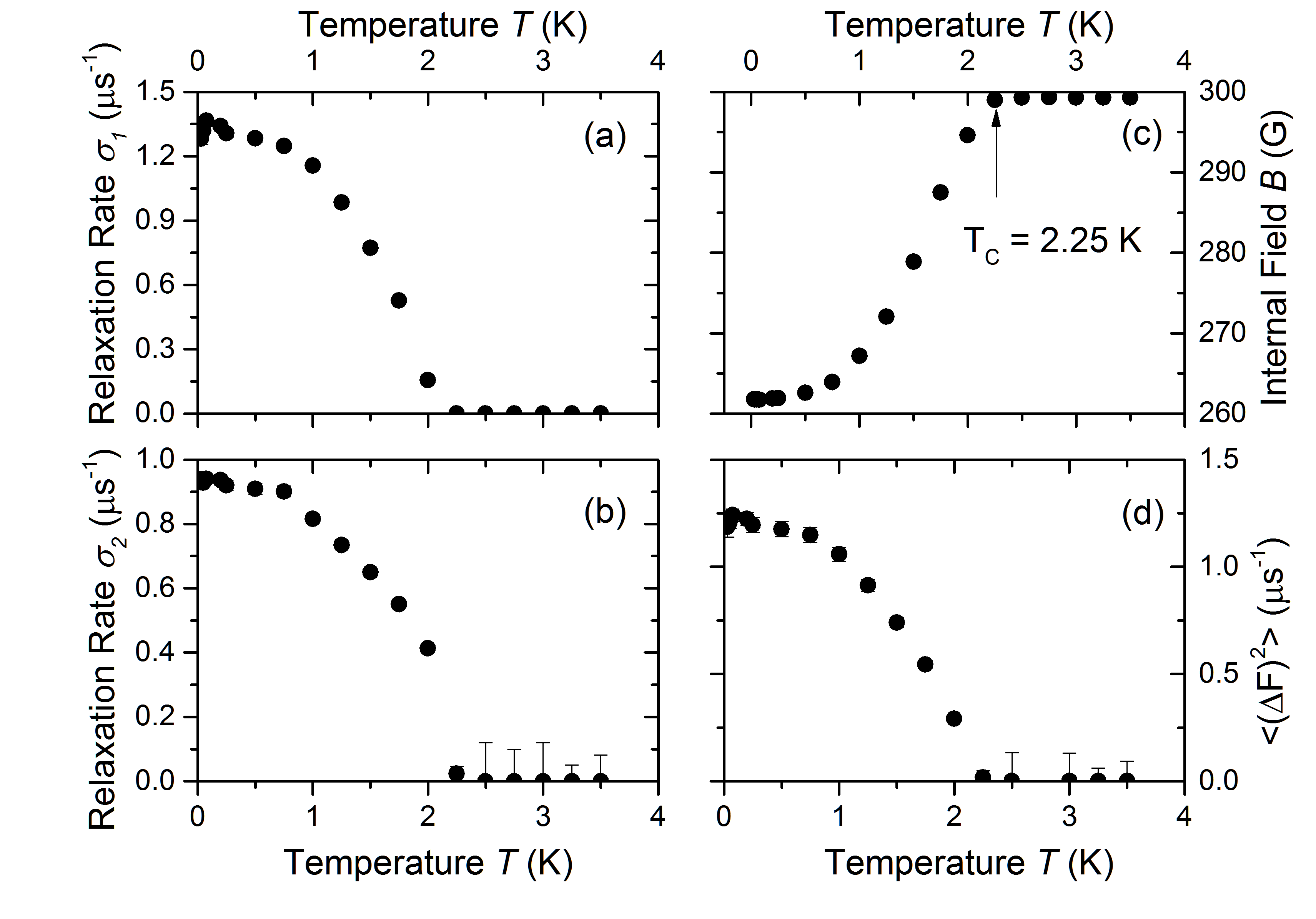}
\caption{Parameters used to fit Eq. \ref{eq:musr} to the $\mu$SR data
 measured in a field of 300~Oe transverse to the muon spins. (a) and (b)
 show the individual relaxation rates $\sigma_1$ and $\sigma_2$. (c) shows
 the sample internal field. (d) shows the central second moment of the fit
 frequency distribution (Eq. \ref{eq:sigma}).}
\label{fig:musrparam}
\end{figure}

\begin{figure}[ht]
\includegraphics[width=\columnwidth]{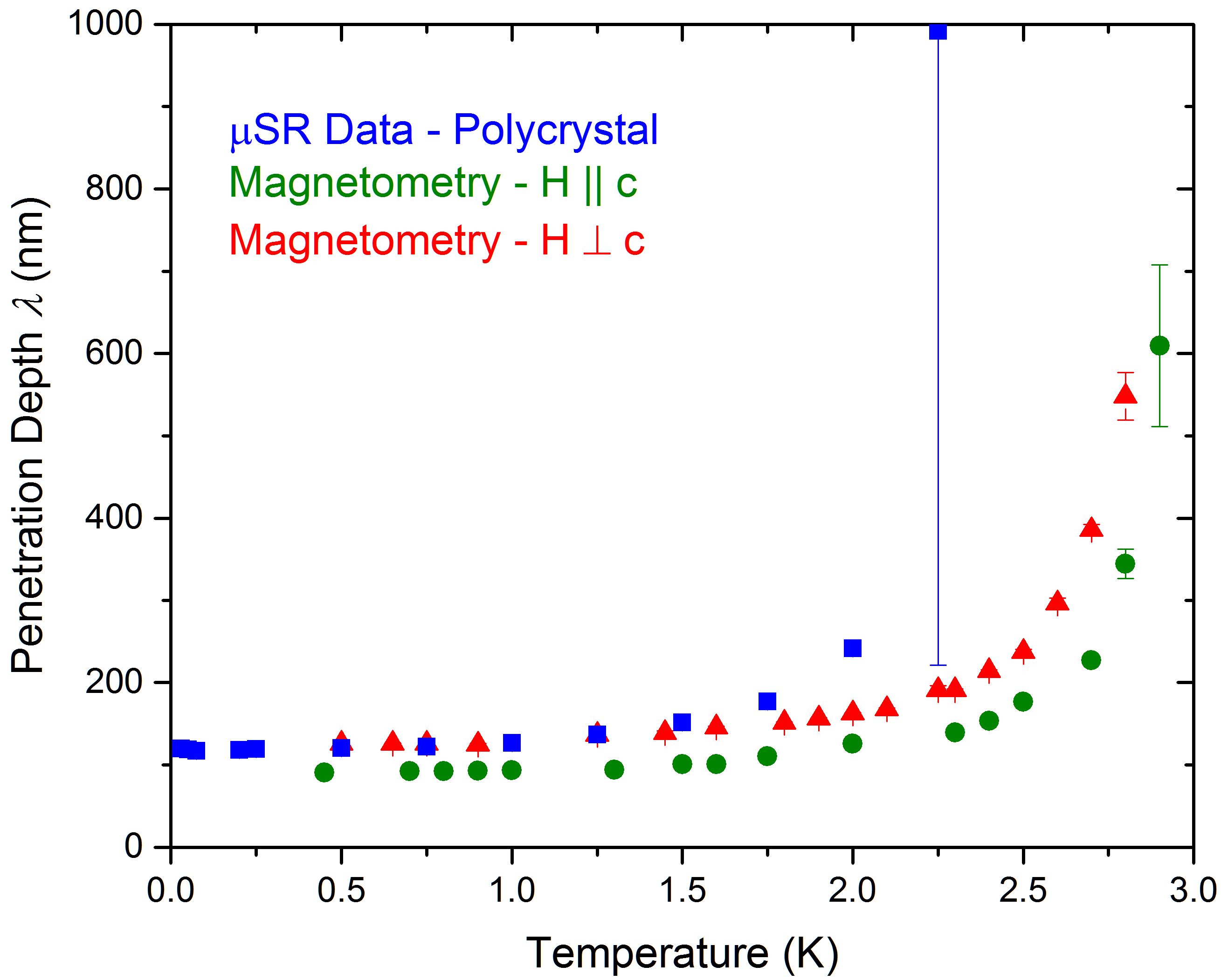}
\caption{Penetration depth determined from magnetometry and $\mu$SR
 measurements. Green circles are from magnetometry of a single crystal
 with H $\parallel$ C-axis. Red triangles are from magnetometry with H
 $\perp$ C-axis. Blue squares are from $\mu$SR using a Gaussian fit.}
\label{fig:PD}
\end{figure}

These fits gave values of $C = 0.7046$, $F = 0.37$, and the temperature
 dependent values shown in Fig. \ref{fig:musrparam}. The temperature
 dependence of the fit parameters indicate that $T_C \approx 2.25$~K,
 consistent with that from our magnetization measurements at the same field.
 From the $\sigma_T$ we calculated the temperature-dependent penetration
 depth using Eq. \ref{eq:lambda}; this is shown in Fig. \ref{fig:PD}
 (blue squares). This penetration depth diverges towards infinity approaching
 $T_C$ and at low temperature (T \textless ~0.5~K) has an average value
 of 119 $\pm$ 2~nm with very weak temperature dependence (linear fit slope
 of 4 $\pm$ 3~nm $\approx$ 0). This behavior is consistent with what is
 expected for a conventional fully gapped superconductor that should asymptote
 to a constant low temperature value. 

To compare with the penetration depth measured by $\mu$SR, we also
 performed magnetization vs. field measurements at a range of temperatures
 below $T_C$ on a single crystal plate. As our field in these measurements
 was applied using a superconducting coil, there will always be some trapped
 flux in the magnet, resulting in an offset from the expected field set by
 applying current. We corrected for this by doing a linear fit of the low-field
 MvH data of the ZFC field scans and subtracting the resulting field offset.
 This indicated a trapped flux of $\approx 2.5$~Oe for the $H$~$||$~C-axis
 measurements, and $\approx7.5$~Oe for $H$~$\perp$~C-axis.

\begin{figure}[ht]
\includegraphics[width=\columnwidth]{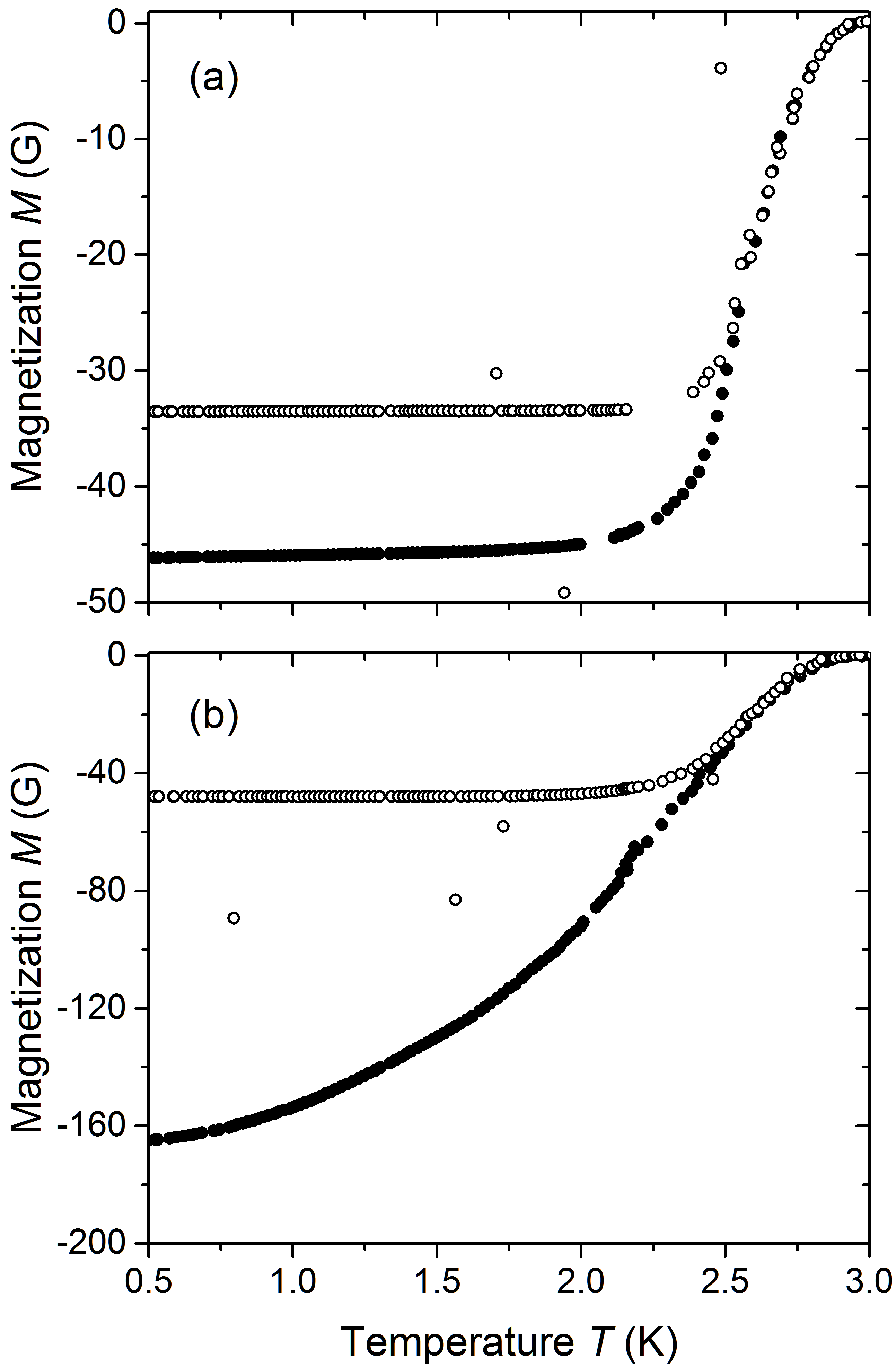}
\caption{Magnetization measurements on a single crystal sample
 of Ir$_{0.95}$Pt$_{0.05}$Te$_2$  in a field of 50~Oe applied (a)
 perpendicular to the C-axis and (b) parallel to the C-axis. Closed
 circles show measurements after cooling in zero applied field and
 open circles show measurements after cooling with the field applied.}
\label{fig:SCMvT}
\end{figure}

Magnetization vs. temperature data for this crystal at 50~Oe
 \textless ~ $H_{c1}$ is shown in Fig. \ref{fig:SCMvT} and
 indicates that $T_C \approx 3$~K at this lower applied field.
 The magnetization in Fig. \ref{fig:SCMvT} (b) is significantly
 larger than 50~G because demagnetization effects increase
 the effective internal field. We accounted for this in the rest
 of the analysis by approximating our sample as a rectangular
 prism of dimensions 2.4~mm x 1.5~mm x 0.35~mm. This
 gives a demagnetization factor of $D_{||} = 0.7039$ for the
 field applied parallel to the C-axis, and $D_{\perp} = 0.1124$
 for the field applied perpendicular to the C-axis, using the
 formula found in Ref. \cite{Aharoni1998}. The internal field
 is then calculated as $H_{int} = H_{ext} - D M$. This gives
 low temperature effective ZFC internal fields of 176~G for $H$ $||$ C-axis,
 and 55~G for $H \perp$ C-axis which indicate that either 98\% or 84\%
 of the volume is superconducting. The discrepancy between these two
 numbers may indicate some inaccuracy in our estimation of the
 demagnetization factors, but this uncertainty does not substantially
 affect the conclusions we have reached. 

The magnetization of a type II superconductor in the reversible
 regime near $H_{c2}$ can be approximated using the London
 model as \cite{Hao1991},

\begin{equation}
-4\pi M = \frac{\alpha\phi_0}{8\pi\lambda_2}\ln\left(
\frac{\beta H_{c2}}{H}\right).
\label{eq:MvHLondon}
\end{equation}

Here, $M$ is the magnetization in G, $\phi_0$ is the flux quantum, $\lambda$
 is the effective zero field penetration depth, $\alpha$ and $\beta$ are
 constants which depend on the field range being fit. We therefore plotted
 $M$ vs. $\ln(H)$, and fit the resulting linear regime to determine $\lambda$
 from the slope ($s$) as,

\begin{equation}
\lambda = \sqrt{\alpha\frac{\phi_0}{8\pi s}}.
\label{eq:slope}
\end{equation}

\begin{figure}[ht]
\includegraphics[width=\columnwidth]{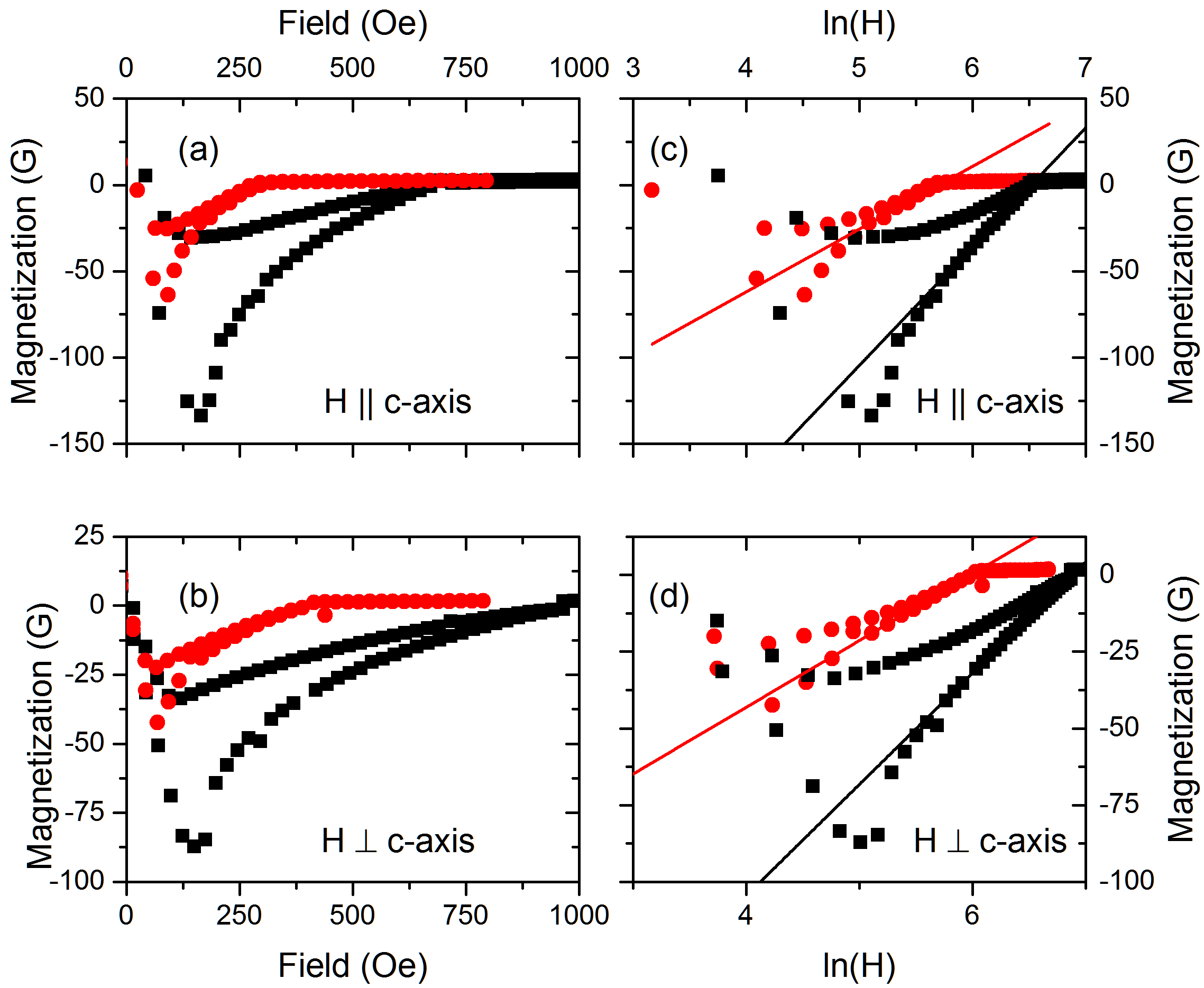}
\caption{(a-b) Magnetization vs. internal field curves measured at 0.5~K
 (black squares) and 2~K (red circles) for (a) H $\parallel$ C-axis and (b)
 H $\perp$ C-axis. (c-d) Magnetization vs. ln(H) curves along with linear
 fits to the high-field region (solid lines) measured at 0.5~K (black squares)
 and 2~K (red circles) for (c) H $\parallel$ C-axis and (b) H $\perp$ C-axis.}
\label{fig:MvH}
\end{figure}

We used an $\alpha$ value of 0.7 in the following analysis, appropriate
 to higher field ranges \cite{Hao1991}. However, it is important to note
 that changing this value will only result in a rescaling of the
 penetration depth; it will not affect the temperature dependence.
 Examples of these linear fits are shown in Fig. \ref{fig:MvH} (c) and (d).
 The resulting penetration depths are plotted alongside that measured by
 $\mu$SR in Fig. \ref{fig:PD} (green circles and red triangles).

This analysis gives low-temperature penetration depths of
 $\lambda_{||}(0) = 91$~nm and $\lambda_{\perp}(0) = 125$~nm,
 which shows that the anisotropy in this material is not large. The low
 temperature penetration depth measured by $\mu$SR (120~nm) falls
 between these two values, which is expected as the polycrystalline $\mu$SR
 sample should result in an averaging of the two penetration depths.
 The $\mu$SR value is closer to the $\lambda_{\perp}$ value, which may
 indicate some preferential orientation of the polycrystalline sample.
 However, as the $\mu$SR data is measured at 300~Oe, we would also
 expect it to have a slightly larger penetration depth compared to the
 effective zero field values from the magnetization fitting. It is thus not
 surprising that the $\mu$SR value is above the average of the two zero-field
 values, and we can say that penetration depths measured by our two
 different techniques seem broadly consistent, giving a true zero field
 average penetration depth close to 100~nm.

From the penetration depth, we determined the normalized superfluid
 density, $n_s$, in each case as,

\begin{equation}
\frac{n_s(T)}{n_s(0)} = \frac{\lambda^2(0)}{\lambda^2(T)}.
\label{eq:sf}
\end{equation}

\begin{figure}[ht]
\includegraphics[width=\columnwidth]{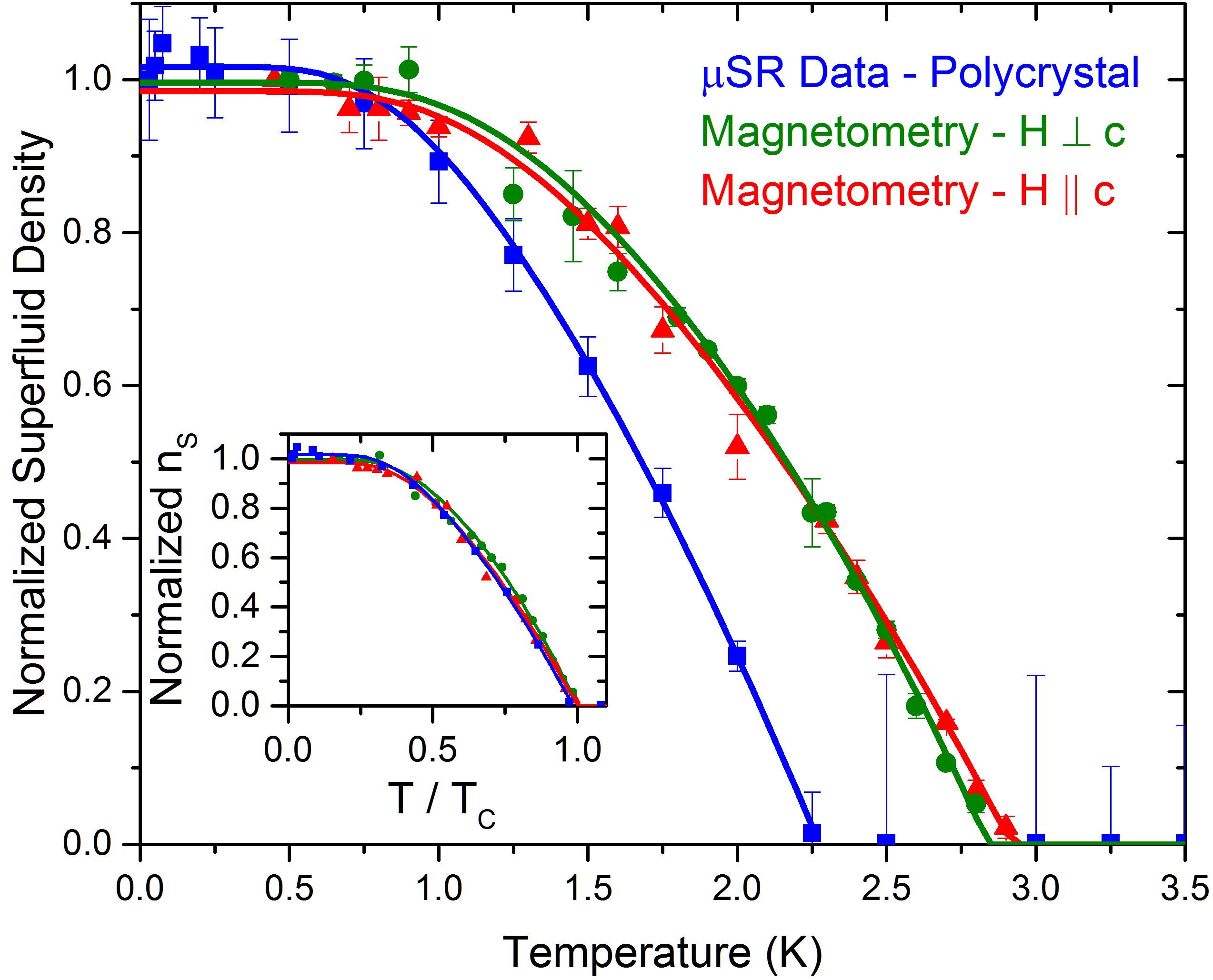}
\caption{Normalized superfluid density determined from magnetization
 and $\mu$SR measurements. Red triangles are from magnetometry
 of a single crystal with H $\parallel$ C-axis. Green circles are from
 magnetometry with H $\perp$ C-axis. Blue squares are from the
 $\mu$SR data. Solid lines show BCS fits to  the data using 
Eq. \ref{eq:superfluid}.}
\label{fig:superfluid}
\end{figure}

The resultant superfluid densities are plotted in Fig. \ref{fig:superfluid}.
 This figure allows us to look at the temperature dependencies of the
 superfluid density in each case without the confounding possible normalization
 issues discussed above. The inset in Fig. \ref{fig:superfluid} shows these
 superfluid densities plotted vs. normalized temperature ($\frac{T}{T_C}$)
 and shows that the temperature dependence of the superfluid density measured
 by the two methods is essentially the same aside from the shift in $T_C$.
 Estimating $H_{c2}$ from our MvH scans gives approximate values of 300~G
 for $H \perp$ C-axis and 225~G for $H \parallel$ C-axis at $T = 2.3$~K, the
 $T_C$ measured from $\mu$SR at 300~G. From these values we would expect
 a somewhat lower $T_C$ at 300~G (closer to 2.1~K), but the discrepancy is not
 large. The likely explanation is that there is some variation between individual
 crystal grains, and that the one we used for the single-crystal measurements
 has a slightly lower $T_C$ compared to the polycrystalline aggregate used for
 the $\mu$SR measurements. 

To determine whether our  data matches what would be expected of a fully
 gapped superconductor, we fit these superfluid densities to the formula
 \cite{Tinkham},

\begin{equation}
n_s(T) = C\left[1-2\int_{\Delta}^{\infty}dE\frac{\partial F}{\partial E}
 \frac{E}{\sqrt{E^2 - \Delta^2}}\right].
\label{eq:superfluid}
\end{equation}

Here, $C$ is a scaling constant, $E$ is the energy difference above the Fermi
 energy, $F = \frac{1}{e^{E/k_BT} + 1}$ is the Fermi function, $k_B$ is the
 Boltzmann constant, and $\Delta$ is the gap, which we approximate using
 the interpolation formula \cite{Gross1986},

\begin{equation}
\Delta(T) = \Delta_0 \tanh\left(1.742 \sqrt{\frac{T_c}{T} - 1}\right).
\label{eq:delta}
\end{equation}

Here, $\Delta_0$ is the zero temperature value of the gap, and $T_c$ is the
 critical temperature.

The results of these fits are shown as the solid lines in Fig.
 \ref{fig:superfluid}. These data all show good agreement with the fits,
 therefore our data is consistent with Ir$_{0.95}$Pt$_{0.05}$Te$_2$
 being a fully gapped superconductor. In particular, the data show a
 flat temperature dependence of $n_s$ at low temperatures, which
 suggests that there are no nodes in the gap and hence the majority
 of the carriers are fully gapped. We find no evidence in these fits for
 unconventional superconductivity, however there are some exotic states
 such as p-wave $k_x \pm ik_y$ that are fully gapped and would be
 indistinguishable from s-wave in our measurements \cite{Luke2000}.

Furthermore, we can compare the fit values for $T_c$ and $\Delta_0$ shown
 in Table \ref{tab:sfit} to the expected constant $\frac{2\Delta_0}{kT_c} = 3.5$
 for a BCS weak coupling superconductor. The data show a range between
 3.57 and 3.9 for this ratio, which is close to the expected ratio. Our data are
 consistent with STM measurements on Ir$_{0.95}$Pd$_{0.05}$Te$_2$ that
 found a value of $\frac{2\Delta_0}{kT_c} = 3.6$\cite{Yu2014}. This indicates
 that differently doped (Pd vs. Pt) IrTe$_2$ display similar superconducting
 properties.

\begin{table}
\begin{tabular}{| c | c | c | c |}
      \hline
 & $\Delta_0$~(meV) & $T_C$~(K) & $\frac{2\Delta}{k_B T_C}$\\
\hline
$\mu SR$ & 0.351 & 2.28 & 3.57 \\
SQUID Perpendicular & 0.463 & 2.84 & 3.90 \\
SQUID Parallel & 0.463 & 2.92 & 3.68 \\
\hline
  \end{tabular}
\caption{Parameters used for the superfluid density fits to Eq. 
\ref{eq:superfluid} shown in Fig. \ref{fig:superfluid}.}
\label{tab:sfit}
\end{table}

\section{Conclusion}
We have presented penetration depth and superfluid density data of 
Ir$_{0.95}$Pt$_{0.05}$Te$_2$ determined from SQUID magnetometry and $\mu$SR.
 These data are consistent with conventional BCS weak coupling s-wave 
superconductivity in Ir$_{0.95}$Pt$_{0.05}$Te$_2$, with a zero temperature gap
 of $\Delta_0 = 0.46$~meV in zero field. The gap decreases to 0.35~meV at
 $H_{ext}$ = 300~G as expected from the corresponding drop in $T_C$ over
 the same field range. We see no evidence for nodes in the gap which suggests
 that d-wave pairing symmetry does not appear in this material. 
However, we are unable to distinguish p-wave and s-wave pairing as some
 p-wave states may be fully gapped.

Finally, our work shows that the temperature dependence of the penetration
 depths measured by two very different techniques ($\mu$SR and magnetometry)
 are consistent with one another. This strengthens the conclusions we can draw
 from one technique alone, and is to our knowledge the first quantitative
 comparison of the results of the two techniques on the same material. 

\section{Acknowledgments}

We thank Dr. G.D. Morris, Dr. B.S. Hitti and Dr. D.J. Arseneau (TRIUMF) for
their assistance with the $\mu$SR measurements. Work at McMaster University was
supported by the Natural Sciences and Engineering Research Council of Canada and
the Canadian Foundation for Innovation. M.N.W acknowledges support from the
Alexander Graham Bell Canada Graduate Scholarship program. 
The Columbia University group acknowledges support from NSF DMR-1436095 (DMREF),
 OISE-0968226 (PIRE), JAEA Reimei project, and Friends of Univ. of Tokyo Inc. 
Work at ORNL was supported by the US Department of Energy, Office of Science,
 Basic Energy Sciences, Materials Sciences and Engineering Division.

\end{document}